\documentclass[twocolumn,groupedaddress,superscriptaddress,lengthcheck]{aastex631}
\usepackage{amsmath}
\usepackage{float}
\usepackage{hyperref}
\usepackage{xspace}

\newcommand{\m}{M\textsubscript{\(\odot\)}\xspace}

\begin{document}

\title{A Search for Low-Mass Neutron Stars in the Third Observing Run of Advanced LIGO and Virgo}

\author[0009-0004-9167-7769]{Keisi Kacanja}
\affiliation{Department of Physics,
Syracuse University, 
Syracuse, NY 13244, USA}

\author[0000-0002-1850-4587]{Alexander H. Nitz}
\affiliation{Department of Physics,
Syracuse University, 
Syracuse, NY 13244, USA}

\correspondingauthor{Keisi Kacanja}
\email{kkacanja@syr.edu}

\begin{abstract}

Most observed neutron stars have masses around 1.4 \m, consistent with current formation mechanisms. To date, no sub-solar mass neutron star has been observed. Observing one would provide crucial constraints on the nuclear equation of state, unveil a new neutron star population, and advance understanding of their formation mechanisms.  We present the first targeted search for tidally deformed sub-solar mass binary neutron stars (BNS), with primary masses ranging from 0.1 to 2 \m and secondary masses from 0.1 to 1 \m, using data from the third observing run of the Advanced LIGO and Virgo gravitational-wave detectors. We account for the tidal deformabilities of up to $O(10^4)$ of these systems, as low-mass neutron stars are more easily distorted by their companion. Previous searches that neglect tidal deformability lose sensitivity to low-mass sources, potentially missing more than $\sim30\%$ of detectable signals from a system with a chirp mass of 0.6 \m  binaries. No statistically significant detections were made. In the absence of a detection, we place a 90\% confidence upper limit on the local merger rate for sub-solar BNSs, constraining it to be  $< 6.4\times10^4$  Gpc$^{-3}$Yr$^{-1}$ for a chirp mass of 0.2 \m and $< 2.2\times 10^3$  Gpc$^{-3}$Yr$^{-1}$ for 0.7 \m. With future upgrades to detector sensitivity, development of next-generation detectors, and ongoing improvements in search pipelines, constraints on the minimum mass of neutron stars will improve, providing the potential to constrain the nuclear equation of state, reveal new insights into neutron star formation channels, and potentially identify new classes of stars.

\end{abstract}

\section{Introduction} \label{sec:intro}
Neutron stars are incredibly dense objects that serve as some of the most unique natural laboratories for exploring the behavior of matter under extreme density and pressure \citep{2018EPJP..133..445V,annurev}. Neutron stars are formed through the cataclysmic events of supernova explosions when massive stars exhaust their nuclear fuel, leading to a collapse under their gravity \citep{2018EPJP..133..445V}. Given the supernova mechanism of generating neutron stars, analytical solutions, and simulations predict that neutron stars typically have masses ranging from 1.2 to 2 \m\citep{2013ApJ...778...66K,2001ApJ...550..426L,Douchin:2001sv,2012ARNPS..62..485L,10.1093/mnras/sty2460}. However, considerable uncertainty remains, especially at the mass range limits and the equation of state of the neutron star core, which warrants further research \citep{2019PhRvD.100h3010F,2024PhRvC.109f5807G}.  

One active area of research focuses on constraining the nuclear equation of state (EOS) of neutron stars. EOS models suggest that a neutron star with a mass of 1 \m could have a radius between 11 and 15 km \citep{2020CQGra..37d5006A,1997A&A...328..274B,MULLER1996508,DANIELEWICZ200936,PhysRevC.72.014310,PhysRevC.53.740}, a 2 \m neutron star may have a radius from 10 to 14 km, and at the extreme ends a 0.4\m neutron star may have radii anywhere from 11 km to 16.5 km. The wide range reflects the significant uncertainty in EOS constraints at the extreme densities, resulting in multiple competing models \citep{2024arXiv240815192B,2024arXiv240711153C,2016ARA&A..54..401O,2018RPPh...81e6902B}. At the upper mass limit, the mass-radius curves of the EOS show that the neutron star is predicted to have a smaller radius than a low-mass neutron star. Extending the EOS to higher masses is difficult due to causality constraints, as the pressure-density relationship would require the speed of sound to exceed the speed of light for masses above 2 \m\citep{2022arXiv220700033S}. At the lower mass limit, some EOS models describe neutron stars as light as 0.1 \m, though few plausible formation mechanisms exist for these stars, and no direct observations have yet confirmed their existence \citep{annurev,2022arXiv221201477T,Nitz_2022,Metzger_2024}. Overall, constraints on the EOS, particularly at the extreme mass limits, still depend on observational data and are expected to improve as more BNS mergers are detected, including those observed through gravitational waves (GWs).

GWs are distortions in spacetime produced by the acceleration of massive objects. These waves can be generated from binary systems of compact objects, such as neutron stars and black holes, where GWs are radiated as the two compact objects inspiral toward each other and eventually merge \citep{2018IJMPA..3330013D}. Ground-based observatories such as  Advanced LIGO and Advanced Virgo measure the tiny distortions caused on the detector by the GWs of the compact binary \citep{2015CQGra..32b4001A,2015CQGra..32g4001L}. To date, two BNS mergers have been detected via GWs \citep{2017PhRvL.119p1101A,2020ApJ...892L...3A}. The first detection, GW170817, had a chirp mass $\mathcal{M}$  of 1.186 \m and was accompanied by an electromagnetic counterpart.\citep{2017ApJ...848L..12A,Alexander:2017aly,LIGOScientific:2017ync,Valenti_2017}. This event confirmed that BNS collisions are the progenitors of kilonovae and gamma-ray bursts \citep{2017PhRvL.119p1101A,LIGOScientific:2017ync}.

GW observations allow for the measurement of the masses and tidal distortions of neutron stars through their imprint on the gravitational waveform. The chirp mass $\mathcal{M}$ is defined as:
\begin{equation} \mathcal{M} = \frac{(m_1m_2)^\frac{3}{5}}{(m_1+m_2)^\frac{1}{5}} \end{equation}
where $m_1$ and $m_2$ are the masses of the two compact objects in the binary. The masses influences the amplitude and frequency of the signal, with larger masses increasing the amplitude of the waveform and shifting the frequency to lower values. Additionally, the phase evolution of the waveform encodes information about the tidal distortions of the stars, often referred to as tidal deformability. Together, the masses and tidal deformability provide direct constraints on the EOS, making GW observations a powerful tool for probing the EOS of neutron stars. Tidal deformability $\lambda$ is defined as 

\begin{equation}
\lambda = \frac{2}{3}k_2\bigg(\frac{Rc^2}{Gm}\bigg)^5    
\end{equation}

where $k_2$ is the tidal love number, $R$ is the neutron star radius, $c$ is the speed of light, $G$ is the gravitational constant and $m$ is the mass of the neutron star.

\begin{figure}
    \hspace*{0.2cm}
    \includegraphics[height=3.5cm,width=7.8cm]{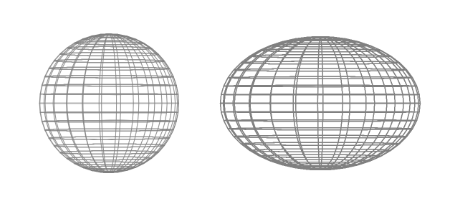}
    \caption{Figure of a non-deformed and tidally deformed neutron star. The left sphere represents a non-tidally deformed neutron star and the right sphere represents a tidally deformed neutron star.}
    \label{fig:cartoon}
\end{figure}

One representation of tidal deformability can be shown in figure \ref{fig:cartoon}.  In the absence of a tidal field, a non-spinning neutron star would be spherical in geometry, whereas a neutron star in a tidal field would be deformed. Smaller mass neutron stars tend to be more easily deformed by tidal effects, as these objects are less dense than their massive counterparts. 

EOS models predict that sub-solar neutron stars can have deformability ranging from $\mathcal{O}(10^3)$ for a 1 \m up to $\mathcal{O}(10^6)$ for a 0.1 \m star \citep{2020CQGra..37d5006A,1997A&A...328..274B,MULLER1996508,DANIELEWICZ200936,PhysRevC.72.014310,PhysRevC.53.740}. The combination of NICER observations and GW data has proven to be a powerful tool for constraining the EOS of neutron stars. In a study by \citep{2021ApJ...918L..29R}, the integration of these observations helped refine the EOS, suggesting a preference for neutron stars with smaller radii $\sim12$km for a 1.4 \m neutron star. However, there are still some uncertainties on the extreme mass limits of the EOS and some of the most stringent constraints on the EOS could come from observations of low-mass binary systems, as the tidal effects in their GW signals are more prominent, providing stronger constraints than those from a 1.4 \m neutron star \citep{2024arXiv240517326K}

\begin{figure*}[t]
    \centering
    \hspace{-0.5cm}
    \includegraphics[height=7cm,width=9cm]{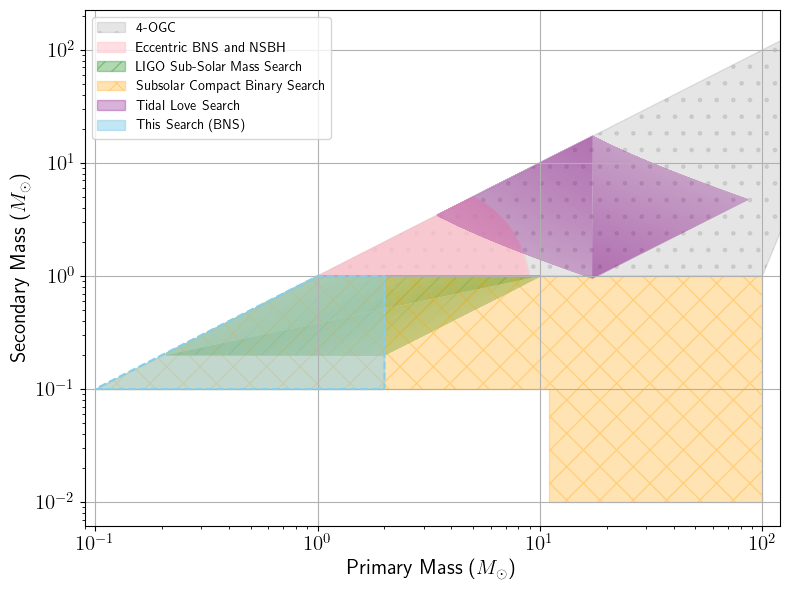}
    \hspace{0.2cm}
    \includegraphics[height=7cm,width=9cm]{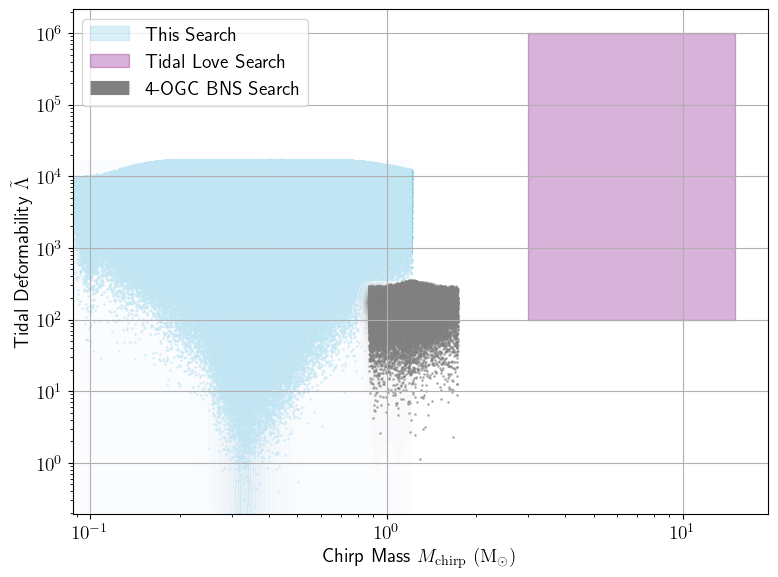}
    \caption{Template Bank parameter spaces for previous searches. The left plot shows the previous searches parameter space with primary mass on the horizontal axis and secondary mass on the vertical axis. The right plot shows all the searches that incorporated tidal deformability versus the chirp mass of the systems. The blue region represents this search considering tidal deformability. The red region corresponds to the bank for eccentric BNSs and Neutron Star Black Hole (NSBH) binaries \citep{dhurkunde2023search}. The yellow bank is the O3 sub-solar mass run considering eccentric and non-eccentric Compact Binaries \citep{Nitz_2022}. The green region represents the sub-solar bank search performed by the LIGO collaboration \citep{2022arXiv221201477T}. The purple region is the tidal love search \citep{Chia:2023tle}. The gray region is the 4-OGC search considering non-eccentric compact binaries \citep{Nitz:2021zwj}.}
    \label{fig:parameter-space}
\end{figure*}

Detecting a low-mass neutron star would provide tighter constraints on the EOS while offering insights into alternative evolutionary pathways for forming ultra-light binaries, and challenging current assumptions about neutron star populations. A recent study by \citealt{2022NatAs...6.1444D} suggests that the supernova remnant HESS J1731-347 could be as light as 0.77 \m. Theoretically, the lowest mass a static neutron star can have is 0.1 \m determined by the stability of the star’s self-gravity \citep{Shapiro:1983du}. Detecting a sub-solar mass event through GWs could not only confirm the existence of low-mass neutron stars but might also point to entirely new types of compact objects, such as quark stars. 

Quark stars, theorized to be composed of quark matter, are expected to have masses ranging from 0.1 to 2 \m making them a plausible candidate for such discoveries \citep{Weber:2012ta, Issifu:2024lgj,2022NatAs...6.1444D, Xu2005}. Unlike traditional neutron stars, quark stars are predicted to exhibit lower tidal deformabilities \citep{2021PhRvD.104h3011A,2024PhRvD.110b3012H}. For example, the tidal deformability of a 1.4 \m neutron star is predicted to be around 500, whereas a quark star of the same mass would have a deformability of about 100, making quark stars less susceptible to tidal distortions by companion objects \citep{2024PhRvD.110b3012H}. This difference grows significantly for lower masses; for example, a 0.5 \m quark star would have a tidal deformability of 600, whereas a neutron star of the same mass would have a tidal deformability on the order of $\mathcal{O}(10^5)$ \citep{2024PhRvD.110b3012H,2020CQGra..37d5006A}. This distinctive difference in deformability could be detectable in gravitational wave signals, providing a potential means to distinguish quark stars from neutron stars.

If the detected object is indeed a neutron star, its low mass suggests it likely formed through alternative mechanisms, as traditional supernovae are not expected to produce such low-mass remnants (no less than 1.2 \m) \citep{2013ApJ...778...66K,2001ApJ...550..426L,Douchin:2001sv,2012ARNPS..62..485L,10.1093/mnras/sty2460,2024arXiv240814287C}. One proposed mechanism for generating a low-mass neutron star involves the fragmentation of the accretion disk around a black hole \citep{Metzger_2024}. In this scenario, the collapse of a rapidly rotating massive star (greater than 20 \m) leads to the formation of a centrifugally supported torus. Matter from the outer layers of the star’s core fragments into multiple pieces, which can then condense into sub-solar mass neutron stars. These systems, which are expected to be in binary configurations, could be detected by ground-based gravitational wave detectors such as Advanced LIGO and Advanced Virgo.

Previous searches have attempted to look for sub-solar mass systems. \citealt{2022arXiv221201477T} and \citealt{Nitz_2022} searched for sub-solar mass binary systems similar to this paper; however, they did not account for the tidal deformability of the system. Not accounting for tidal deformability for low-mass sources results in a loss of up to 78.4\% of the signals for an equal 0.2 \m system \citep{2023PhRvD.107j3012B}. This loss was calculated under idealized conditions by comparing matches between templates and fiducial signals. In practice, gravitational wave signals are buried in noise, making the actual sensitivity loss likely to be higher. Thus, the 78.4\% loss represents an optimistic estimate, as further reductions in sensitivity are expected due to signal-template mismatches introduced by signal-consistency tests in the search pipeline \citep{Usman:2015kfa,2017PhRvD..95d2001M}. Consequently, if a nearby sub-solar BNS system had merged, the signal might have been missed by previous searches due to reduced sensitivity from not accounting for tidal deformability. Figure \ref{fig:parameter-space} shows the parameter space explored in previous searches.  Our search cover a significant portion of the effective tidal deformability $\tilde{\Lambda}$ for low mass systems.  $\tilde{\Lambda}$ is defined as
\begin{equation}
\tilde{\Lambda}  =  \frac{16}{13} \frac{(12q+1)\lambda_1 +(12+q)q^4\lambda_2}{(1+q)^5}    
\end{equation}
where $q = m_2/m_1$ where $m_1>m_2$, and $\lambda_1,m_1$ and $\lambda_2,m_2$ are for each respective neutron star in the binary. No previous sub-solar mass search has incorporated effective tidal deformability. To account for this loss of sensitive volume, we performed a novel search to explicitly look for highly deformable sub-solar BNS systems to ensure none of the low-mass neutron stars have escaped detection by previous search criteria. We perform a search using the public data from the third observing run (O3) of advanced LIGO and Advanced Virgo detectors. We detect no new mergers but use our observations to place an upper limit on the merger rate. 

\section{Search} \label{sec:population}

We utilized the \textsc{PyCBC} gravitational wave analysis software to perform a template-based matched filtering search, aiming to identify sub-solar mass neutron stars in the GW data \citep{Nitz:2017svb,Usman:2015kfa}. Matched filtering is performed by comparing the data to a set of modeled waveforms, known as a template bank, and identifying peaks in the signal-to-noise ratio (SNR). These peaks, or triggers, are checked for coincidences between different detectors in the network, and the resulting coincidences are used to calculate false alarm rates and assign statistical significance to candidate events. To evaluate the detectability of a source, a ranking statistic is computed and assigned for each candidate. The ranking statistic evaluates candidates by estimating the probability that a candidate arises from noise fluctuations, rather than from a genuine astrophysical signal \citep{2017ApJ...849..118N}. The larger the statistic the more likely the candidate is associated with an astrophysical origin.  To make a confident detection, a candidate must be assessed based on a sufficiently high network SNR, a large ranking statistic, and low false alarm rates \citep{PhysRevLett.116.061102}.

To conduct a modeled search, a template bank is required to optimally place waveforms for efficient SNR recovery \citep{2012PhRvD..85l2006A,PhysRevD.44.3819,PhysRevD.53.6749,PhysRevD.60.022002,1996PhRvD..53.3033B,2009PhRvD..80j4014H,2012PhRvD..86h4017B,2014PhRvD..90h2004D,PhysRevD.99.024048,PhysRevD.108.042003}. We constructed our template bank using the stochastic placement method, as described in \citep{2024ApJ...975..212K}, which allows flexibility in choosing any N-dimensional sets of parameters. Current geometric template banks are unable to incorporate tidal deformability because constructing a metric that accurately accounts for tidal effects is highly challenging \citep{2021PhRvD.104d3008H}. As a result, we employed a stochastic method, which allowed us to flexibly generate waveform models with significant tidal deformabilities. The GW signals in the analysis were modeled using the \textsc{TaylorF2} approximant \citep{Droz:1999qx,Blanchet:2002av,Faye:2012we}. 

\begin{table}[ht]
\centering
\hspace{-2cm}
\begin{tabular}{|l|l|p{8cm}|}
\hline
\textbf{Parameter} & \textbf{Range}  \\ \hline
Primary Mass \( m_1 \) & \( [0.1, 2] \, M_\odot \) \\ \hline
Secondary Mass \( m_2 \) & \( [0.1, 1] \, M_\odot \) \\ \hline
Chirp Mass \(\mathcal{M}\) & \( [0.087, 1.22] \, M_\odot \) \\ \hline
Aligned Spin \( \chi_1, \chi_2 \) & \( [-0.05, 0.05] \) \\ \hline
Tidal Deformability \( \lambda_1, \lambda_2 \) & \( [0, 10^4] \) \\ \hline
Effective Tidal Deformability \( \tilde{\Lambda} \) & \( [0.19, 1.7\times 10^4] \) \\ \hline

\end{tabular}
\caption{Template bank parameter space. The template duration was fixed to 512 seconds and was constructed to recover 95 \% of the SNR.}
\label{tab:template_bank}
\end{table}

Our template bank covers the parameter ranges depicted in Table \ref{tab:template_bank}. We choose the component tidal deformabilities to go up to $10^4$ to account for the EOS of low-mass neutron stars \citep{2020CQGra..37d5006A,1997A&A...328..274B,MULLER1996508,DANIELEWICZ200936,PhysRevC.72.014310,PhysRevC.53.740}. Extending the analysis to cover tidal deformability up to $\mathcal{O}(10^6)$ was not feasible due to the rapid growth in template bank size, which would make the search computationally prohibitive. 
The greatest sensitivity loss from tidal deformability reported in \citep{2023PhRvD.107j3012B} occurs for the lowest-mass system they considered. For instance, a binary with a chirp mass of 0.14 \m would lose more than 80\% of the number of signals, while a binary with a chirp mass of 0.6 \m loses approximately 27\% of the volume. Unequal-mass systems, such as a 0.2–1 \m binary, experience a loss of more than 27\% of the signals. To enhance sensitivity across the lower mass companions, our template bank includes unequal binaries with primary masses up to 2 \m, extending down to the theoretical stable limit of 0.1 \m, and secondary masses ranging from 0.1 to 1 \m. 
We assume the spins of these systems to be small, as there are no known formation channels that suggest sub-solar mass neutron stars would be rapidly spinning. Most observed galactic binary neutron stars have spins $<0.05$ \citep{2017ApJ...846..170T}, and thus we choose the spin parameters to be $\chi_1, \chi_2 \in [-0.05, 0.05]$.

Stochastic template bank methods do not guarantee recovery of all possible signals at or above the minimal match threshold, potentially leaving gaps in coverage. We validated the bank’s completeness by randomly generating waveform models and calculating matches between injected signals and their nearest templates. The bank was designed such that 99\% of templates successfully recover SNRs $\geq 95\%$ for fiducial injections within the parameter space. However, coverage was weakest for equal-mass systems at 0.1 \m. Less than 1\% of the bank failed to recover matches $\geq 0.95$. The final bank contained approximately $10^7$ templates. We also tested the bank with a more modern waveform model to evaluate any potential improvement in sensitivity. We compared the recovered SNRs from injections generated with \textsc{IMRPhenomXP\_NRTidalv2} to those from our final bank. We found that the recovered SNRs were consistent with the \textsc{TaylorF2} injections, showing no improvement in sensitivity.

Omitting tidal deformability from the bank would result in an $O(10^1)$ factor reduction in templates \citep{2024ApJ...975..212K}. For comparison, the template bank used in the low-mass neutron star detectability study by \citep{2023PhRvD.107j3012B} comprised approximately $1.7 \times 10^6$ templates, covering equal mass components $m_1, m_2 \in [0.2, 1]$ \m but also a wider spin range up to 0.1. The sub-solar mass search conducted by \citep{Nitz_2022} employed a bank with $7.8 \times 10^6$ templates, accommodating mass ranges $m_1 \in [0.1, 7]$ \m and $m_2 \in [0.1, 1]$ \m as well as eccentricity $e \in [0, 0.3]$. This search utilized the largest template bank to date and was the most computationally demanding, requiring approximately $2.2 \times 10^8$ core days to complete.

\section{Results} \label{sec:results}
The search analyzed data from the O3a and O3b observing runs of Advanced LIGO and Advanced Virgo, spanning April 1, 2019, 15:00 UTC (GPS 1238166018) to October 1, 2019, 15:00 UTC for O3a, and November 1, 2019, 15:00 UTC to March 27, 2020, 17:00 UTC for O3b \citep{KAGRA:2023pio}.

No detections of sub-solar mass neutron stars were made. Our best two candidates are shown in table \ref{tab:candidates}. One candidate of a 1.46-0.24 \m system has a rate of 7.14 false alarms per year. The ranking statistic of this source was 9.98 and the network SNR was 9.59. The second candidate was a 0.27-0.1 \m system with a network SNR of 9.18, a ranking statistic of 13.91, and a false alarm rate of 13.91. 

\begin{table}[ht]
    \hspace{-1.7cm}
    \centering
    \resizebox{0.83\linewidth}{!}{ 
        \begin{minipage}{\linewidth}
            \centering
            \begin{tabular}{p{1.0cm} *{5}{>{\centering\arraybackslash}p{1.5cm}}} 
                \hline
                GPS Time & $\mathcal{M}$ [$M_\odot$] & $\tilde{\Lambda}$ & FAR & Ranking Statistic & Network SNR \\
                \toprule
                1247797722 & 0.48  & $1.2\times10^4$ & $7.14$ & 9.98 & 9.59\\
                \hline
                1244168859 & 0.14 & $8.7\times10^3$ & $9.09$ & 13.91 & 9.18\\
                \hline
            \end{tabular}
            \label{tab:candidates}
        \end{minipage}
    }
    \caption{Table of top 2 candidates given the intrinsic source parameters, merger time, ranking statistic, false alarm rate, and the network SNR} 

\end{table}

From our missed detection criteria we can calculate the upper limit of the merger rate $\mathcal{R}(\theta,z)$ for a given astrophysical model. To constrain the upper limit of the merger rate given our set of populations, the expected number of detected events $N$ is given by \citep{2009CQGra..26q5009B,2016ApJ...833L...1A}

\begin{equation}
     N = \int dt \int d\theta dz  \frac{dV}{dz} \frac{d\mathcal{R}(\theta,z)}{d\theta} \frac{\epsilon(\theta, z)}{(1+z)}  
\end{equation}

Where  $t$ is the period of observing time, $(dV/dz)$ is the differential commoving volume, $d\mathcal{R}(\theta,\zeta,z)/d(\theta)$ is the differential merger rate density, $\epsilon(\theta,z)$ is the detection efficiency, which is the probability of measuring a detection within a given set of parameters $\theta$, and redshift z of the binary. Since our search is only sensitive to nearby sources, we do not consider the cosmological effects. The number of detected events will be reduced to

\begin{equation}
     N = \int dtdV\int d\theta \frac{d\mathcal{R}(\theta)}{d\theta} \epsilon(\theta) 
\end{equation}

Note that in this case, $\epsilon$ still depends on redshift and $z$ is absorbed into $\theta$. For a given set of intrinsic parameter $(\theta_0)$ we can integrate out the rate density

\begin{equation}
     N = {R}(\theta_0)\int dtdVd\theta \epsilon(\theta) 
\end{equation}
where ${\langle VT\rangle}_{\theta_0} = \int dtdVd\theta \epsilon(\theta)$

Given that no detections were made, we can calculate an upper limit of the merger rate $\mathcal{R}_{90}$ using
\begin{equation}
     \mathcal{R}_{90}= \frac{2.303}{\langle VT\rangle}
\end{equation}

Where 2.303 is the expected number of detections $N$ given a 90\% confidence drawn from a Poisson distribution \citep{2009CQGra..26q5009B}.
\begin{figure}
    \hspace{-0.3cm}
    \includegraphics[height=7.5cm,width=8.6cm]{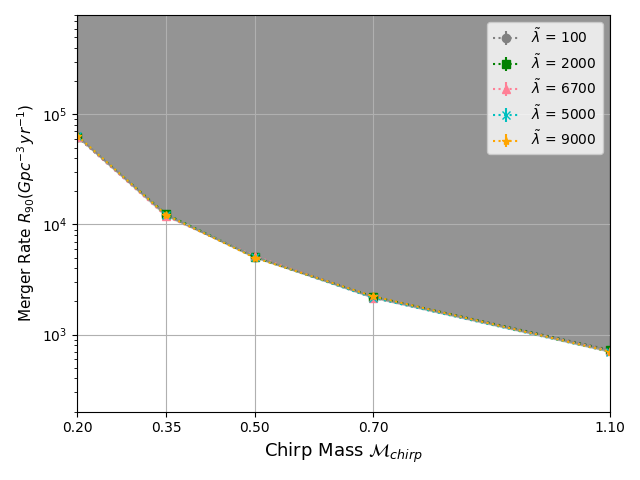}
    \caption{Upper limit of the merger rate versus different chirp mass distribution. The search is equally sensitive across all tidal deformability.}
    \label{fig:merger-rate}
\end{figure} 

We calculated the averaged volume time product $\langle VT\rangle$ using a Monte Carlo integral method for the injected population of signals by counting the number of missed and found injections from our search pipeline. We injected the signals assuming an isotropic distribution, with sources randomly distributed across sky positions and orientations. We fixed five chirp masses and five tidal deformabilities. Figure \ref{fig:merger-rate} shows the upper limit for these fixed parameters. Our search is equally sensitive across all tidal deformabilities. Overall, we find that our search constrains the upper limit of the merger rate for binary neutron stars to be $< 6.4\times 10^4$ Gpc$^{-3}$Yr$^{-1}$ for chirp mass 0.2 \m and $< 2.2\times10^3$ Gpc$^{-3}$Yr$^{-1}$ for 0.7 \m. We find that our results are consistent with previous non-tidally deformed sub-solar mass searches. For a 1.1 chirp mass system, previous searches rates have calculated the upper limit to be $\sim7\times 10^2$ \citep{2022arXiv221201477T}, where we have  $< 7\times 10^2$ Gpc$^{-3}$Yr$^{-1}$ for all O3 data and tidally deformed sources.

\section{Discussions and Conclusion} \label{sec:conclusion}

In this study, we first constructed a template bank for sub-solar mass systems accounting for tidal deformability and spin. Next, we conducted a dedicated search for low-mass neutron star binaries over the third observing run of LIGO and Virgo. We did not discover any potential candidates with a false alarm rate lower than 1 per year. To constrain the upper limit of the merger rate, we assess the sensitivity of our search by simulating a population of low-mass neutron stars with fixed chirp mass and tidal deformability and identify if the search detected or missed the fiducial signals. Our results constrain the merger rate to be $< 6.4\times10^4$ Gpc$^{-3}$Yr$^{-1}$ for a chirp mass of 0.2 \m and $< 2.2\times10^3$ Gpc$^{-3}$Yr$^{-1}$ for 0.7 \m.

Detecting a low-mass neutron star through GWs would represent a breakthrough in astrophysics, challenging current theories on neutron star formation, stellar evolution, and the EOS of neutron stars. Such a discovery could support alternative formation mechanisms, such as fragmentation in accretion disks of massive progenitors \citep{Metzger_2024}, or uncover new formation channels. It would also necessitate a reassessment of the neutron star population distribution, suggesting a previously undetected class of low-mass compact objects. Furthermore, a low-mass system detection would place tighter constraints on the EOS since the larger tidal deformability would make the EOS more readily distinguishable from noise \citep{2024arXiv240517326K}.

This discovery could also drive the investigation of exotic matter states, such as quark matter and dark matter, potentially revealing unique material properties under extreme densities and pressures. If an exotic star, such as a quark star, can be distinguished from a neutron star, it may offer new opportunities for detecting dark matter signatures \citep{2024PhRvD.110b3012H}. Previous research suggests that a quark star's strong gravitational fields might capture dark matter \citep{2024PhRvD.110b3012H}. Observing GWs from low-mass stars that interact with dark matter could offer a novel way to measure dark matter directly, as such stars might contain dark matter components that influence their structure and tidal deformability. Conversely, if a sub-solar object is found to have no tidal deformability, it could be distinguished from stars as a primordial black hole, another intriguing theoretical candidate that may host clouds of dark matter, providing an alternative means of probing dark matter properties  \citep{2024PhRvD.109l4063C}.

This search provided several insights for computational challenges for future low-mass searches. One of the main limiting factors is the number of templates used. Our bank covered the potential tidal deformable range of up to $\mathcal{O}(10^4)$ for a 0.1 \m neutron star. Extending the analysis to cover the full tidal deformability range (up to $\mathcal{O}(10^6)$) was not feasible due to the rapid growth of the template bank. Our search was also restricted to the most sensitive publicly available data. A recent study suggests that data from the fourth and fifth observing runs of LIGO, Virgo and KAGRA could have sufficient sensitivity to detect sub-solar mass events, should they exist \citep{2024PhRvD.109l4063C,2024arXiv240814287C}. Future searches may revisit this search by incorporating the full tidal deformability range and performing the search over all the current and future available data. To accomplish this in a reasonable amount of time, future searches will require new methodologies to handle larger, more complex parameter spaces efficiently. 

Efforts are already underway to improve the sensitivity and efficiency of modeled searches. For instance, hierarchical search methods are a new efficient method for performing modeled searches. As demonstrated by \citet{Soni:2024cdb}, these methods have been shown to increase the sensitive volume by 10-20\% or the SNR by approximately 6\% while reducing computational costs by a factor of 2.5 compared to the traditional approach used in our work. \citep{2022PhRvD.105j3001D} also employed a hierarchical approach and has shown to increase computational efficiency by a factor of $\sim 10$ for sources with SNR of 6 and $\sim 5$ for templates with SNR of at least 5. Faster methods will make it feasible to explore the full search space, covering the entire range of tidal deformability and utilizing all available data to maximize detection capabilities.

Since the merger rate for these systems is low, it may be unlikely to have nearby low-mass BNSs and the sensitivity of current GW detectors may not yet be sufficient to capture the gravitational strain emitted by such low-mass binaries. With the advent of next-generation observatories, like the Einstein Telescope \citep{2010CQGra..27a5003H,2010CQGra..27s4002P} and Cosmic Explorer \citep{2019BAAS...51g..35R,2023arXiv230613745E}, the detection prospects will improve dramatically. For example, a 40 km Cosmic explorer could potentially see up to a redshift of 10 for a system with a total mass of 2 \m \citep{2023arXiv230613745E}, enabling a more sensitive exploration and reach to more NS populations. The advent of the next generation of GW detectors will provide a means for understanding the densest states of matter, the role of dark matter in compact stars, discovering potential low-mass remnants, and unifying the EOS of neutron stars.

All the necessary files used for this search is available in this \href{https://github.com/kkacanja/Low_Mass_BNS_Search.git} {github repository}.

\section*{Acknowledgments}
KK and AHN acknowledge support from NSF grant PHY-2309240. KK and AHN acknowledge the support from Syracuse University for providing the computational resources through the OrangeGrid High Throughput Computing (HTC) cluster supported by the NSF award ACI-1341006. KK acknowledges the support from the Open Science Grid (OSG) and the Atlas cluster computing team at Albert-Einstein Institute (AEI) Hannover for providing additional computational resources to perform the search.  Additionally, KK would like to express her gratitude to Kanchan Soni, Rahul Dhurkunde, Ananya Bandopadhyay, and Aleyna Akyüz for all the discussions and feedback received.

\bibliography{sample631}{}
\bibliographystyle{aasjournal}

\end{document}